\documentclass[aps,prc,twocolumn,superscriptaddress]{revtex4-1}
\usepackage{graphicx}
\usepackage{amsmath}
\usepackage{braket}
\usepackage{url}
\usepackage{ulem}
\usepackage{multirow}
\usepackage{amssymb} 
\usepackage{booktabs}
\usepackage{rotating}
\usepackage{bm}
\usepackage{bbm}
\usepackage{longtable}
\usepackage{subfigure}
\usepackage{tikz}
\usetikzlibrary{matrix}

\usepackage[colorlinks,
linkcolor=blue,
anchorcolor=red,
urlcolor=red,
citecolor=red]{hyperref}

\begin{document}

\title{\textit{Ab initio} valence-space in-medium similarity renormalization group calculations for neutron-rich P, Cl, and K isotopes}

\author{M. R. Xie}
\affiliation{Institute of Modern Physics, Chinese
Academy of Sciences, Lanzhou 730000, China}
\affiliation{School of Nuclear Science and Technology, University of Chinese Academy of Sciences, Beijing 100049, China}

\author{L. Y. Shen}
\affiliation{Institute of Modern Physics, Chinese
Academy of Sciences, Lanzhou 730000, China}
\affiliation{School of Nuclear Science and Technology, University of Chinese Academy of Sciences, Beijing 100049, China}

\author{J. G. Li}\email[]{jianguo\_li@impcas.ac.cn}
\affiliation{Institute of Modern Physics, Chinese
Academy of Sciences, Lanzhou 730000, China}
\affiliation{School of Nuclear Science and Technology, University of Chinese Academy of Sciences, Beijing 100049, China}

\author{H. H. Li}
\affiliation{Institute of Modern Physics, Chinese
Academy of Sciences, Lanzhou 730000, China}
\affiliation{School of Nuclear Science and Technology, University of Chinese Academy of Sciences, Beijing 100049, China}

\author{Q. Yuan}
\affiliation{Institute of Modern Physics, Chinese
Academy of Sciences, Lanzhou 730000, China}
\affiliation{School of Nuclear Science and Technology, University of Chinese Academy of Sciences, Beijing 100049, China}

\author{W. Zuo}
\affiliation{Institute of Modern Physics, Chinese
Academy of Sciences, Lanzhou 730000, China}
\affiliation{School of Nuclear Science and Technology, University of Chinese Academy of Sciences, Beijing 100049, China}

%\email[]{nicolas.michel@impcas.ac.cn}

\date{\today}

\begin{abstract}
Neutron-rich P, Cl, and K isotopes, particularly those with neutron numbers around $N=28$, have attracted extensive experimental and theoretical interest.
We utilize the \textit{ab initio} valence-space in-medium similarity renormalization group approach, based on chiral nucleon-nucleon and three-nucleon forces, to investigate the exotic properties of these isotopes.
Systematic calculations of the low-lying spectra are performed. 
A key finding is the level inversion between $3/2_1^+$ and $1/2_1^+$ states in odd-$A$ isotopes, attributed to the inversion of $\pi 0d_{3/2}$ and $\pi 1s_{1/2}$ single-particle states.
\textit{Ab initio} calculations, which incorporate the three-nucleon forces, correlate closely with existing experimental data.
Further calculations of effective proton single-particle energies provide deeper insights into the shell evolution for $Z=14$ and $16$ sub-shells.
Our results indicate that the three-body force plays important roles in the shell evolution for $Z=14$ and $16$ sub-shells with neutron numbers ranging from 20 to 28.
Additionally, systematic \textit{ab initio} calculations are conducted for the low-lying spectra of odd-odd nuclei. 
The results align with experimental data and provide new insights for future research into these isotopes, up to and beyond the drip line.

\end{abstract}


\maketitle

\section{Introduction}
An atomic nucleus represents a complex quantum many-body system, where neutrons and protons are bound together through strong nuclear interaction. Our comprehension of nuclear structure is fundamentally anchored in the shell model, a profound theoretical framework that delineates how nucleons are arranged within the nucleus~\cite{PhysRev.75.1969, PhysRev.75.1766.2}. 
The advent of rare isotope beams has revealed that properties of nuclei with extreme $N/Z$ ratios within the nuclear chart strongly modify to the well-known shell structure~\cite{BROWN2001517, SORLIN2008602, GADE2008161, RevModPhys.92.015002}. 
Systematic explorations into exotic nuclei have unveiled that traditional shell gaps are not immutable; rather, they evolve, leading to the disappearance of traditional magic numbers and the emergence of new shell closures~\cite{RevModPhys.92.015002, PhysRevC.81.064328, Wienholtz2013, PhysRevLett.120.062503, LI2023137893, LI2023138197}. 
Significant shell gaps observed in $^{22}$O and $^{24}$O suggest new magic numbers at $N=14,16$~\cite{PhysRevC.92.034316,PhysRevLett.96.012501,LI2023138197}, respectively, with further discoveries of magic numbers at $N=32,34$ in Ca isotopes~\cite{wienholtz2013masses,Steppenbeck2013,PhysRevC.102.034302}.
Additionally, the halo phenomenon in $^{11}$Li suggest the disappearance of $N=8$ shell closure, and studies on isotopes such as $^{29}$F, $^{39}$Na, and $^{40}$Mg indicate the disappearance of magic numbers at $N=20$ and $28$~\cite{PhysRevLett.112.142501, PhysRevC.81.021304, PhysRevLett.124.222504, PhysRevC.101.031301, PhysRevLett.122.052501,PhysRevC.90.014302,PhysRevLett.111.212502}.
These discoveries have not only profoundly propelled our understanding of nuclear physics but also challenged pre-existing paradigms of nuclear stability and structure.

Alterations in nuclear shell structures, primarily driven by monopole interaction, are a focal point in contemporary nuclear physics research~\cite{PhysRevLett.95.232502, PhysRevLett.104.012501, PhysRevC.86.034314,PhysRevLett.114.202501}.
A significant focus of both experimental and theoretical studies has been the monopole shift observed in proton-hole states within neutron-rich isotopes~\cite{Fridmann2005, SUN2020135215, PhysRevLett.110.172503}.
In K isotopes, ranging from $^{39}$K to $^{47}$K, a noteworthy shift in the energy levels between the $1/2^+_1$ and $3/2^+_1$ states is observed, culminating in an inversion at $^{47}$K where the $1/2^+_1$ state becomes the ground state~\cite{BJERREGAARD1967568, NEWMAN1968366, TOUCHARD1982169}. 
This inversion, particularly evident at  $N=28$, signifies an energy level reversal between the $\pi 0d_{3/2}$ and $\pi 1s_{1/2}$ orbits and highlights the crucial role of tensor force in this isotopes~\cite{PhysRevC.79.014310, RevModPhys.77.427, PhysRevLett.95.232502}. 
Interestingly, this inversion is not limited to K isotopes. A similar inversion between the $1/2^+_1$ and $3/2^+_1$ states also occurs in the Cl isotopic chain~\cite{PhysRevC.86.024321, PhysRevC.104.044331}.
This raises the question: what happens after $N=28$ for more rich neutrons?
Furthermore, recent advancements using $\gamma$-ray spectroscopy and exclusive parallel momentum distribution analysis have established a ground state spin $3/2^+$ for $^{51}$K and $^{53}$K, suggesting that the order of the $\pi 0d_{3/2}$ and $\pi 1s_{1/2}$ orbits returns to normal in the vicinity of $N=32$ and $34$~\cite{SUN2020135215, PhysRevLett.110.172503}. 
Nevertheless, the precise dynamics of energy splitting in neutron-rich P, Cl, and K isotopes close and beyond the drip line remain an area for further investigation. 

Understanding neutron-rich isotopes of P, Cl, and K is constrained by limited and occasionally contradictory data. A case in point is $^{49}$K, characterized by an extensive level scheme with all spins, including the ground state, remaining tentative~\cite{PhysRevC.82.034319}. 
Theoretical efforts to describe nuclei in this region have faced challenges.
Notably, large-scale shell model calculations using the SDPF-U phenomenological effective interaction~\cite{PhysRevC.79.014310} have been employed to describe the properties of those nuclei. 
Additionally, the \textit{ab initio} Gorkov self-consistent Green's function method has contributed to understanding neutron-rich Cl and K isotopes~\cite{PhysRevC.104.044331, PhysRevC.101.014318, SUN2020135215}. However, only the $1/2^+_1$ and $3/2^+_1$ states are mentioned~\cite{PhysRevC.104.044331,PhysRevC.101.014318,SUN2020135215}.
The \textit{ab initio} valence-space in-medium similarity renormalization group (VS-IMSRG) approach, based on chiral nucleon-nucleon ($NN$) and three-nucleon ($3N$) interactions, provides a systematic and reliable prediction of nuclear properties~\cite{HERGERT2016165,doi:10.1146/annurev-nucl-101917-021120, LI2023138197, PhysRevC.107.014302, YUAN2024138331}. 
The VS-IMSRG have been successfully adopted to describe the properties of nuclei within this region, such as the unique deformation properties of $^{40}$Mg and $^{42}$Si~\cite{YUAN2024138331}. 
Furthermore, VS-IMSRG calculation has illuminated the emergence of the $N=32$ and $34$ sub-shell closures, particularly in calcium chain~\cite{PhysRevLett.121.022506, PhysRevC.99.064303, PhysRevLett.120.062503, PhysRevLett.120.232501}.

This paper is structured as follows: Firstly, we outline the foundational principles of the \textit{ab initio} VS-IMSRG theoretical framework. 
We then delve into a comprehensive analysis of the low-lying spectra for odd-$A$ neutron-rich P, Cl, and K isotopes, employing two sets of chiral $NN+3N$ interactions. 
This analysis illuminates the shell evolution across these isotopic chains for neutron numbers ranging from $N=22$ to 40, utilizing effective single-particle energies (ESPEs) to elucidate nuclear shell evolutions.
Subsequently, the spectra of these neutron-rich isotopes of odd-odd nuclei were predicted with our results open to future experimental validation. Finally, a summary of the article is given.

\section{Method}
The intrinsic Hamiltonian for an $A$-nucleon system is formulated as 
\begin{equation}
H=\sum_{i<j}^{A}\left(\frac{(\boldsymbol{p}_{i} - \boldsymbol{p}_{j})^2}{2m A} + v_{i j}^{\mathrm{NN}}\right)+\sum_{i<j<k}^{A} v_{i j k}^{3 \mathrm{N}},
    \label{H_in}
\end{equation}
where $\boldsymbol{p}$ denotes the momentum of nucleons in the laboratory, $m$ refers to the nucleon mass, and $v^{\rm NN}$ and $v^{\rm 3N}$ represent the $NN$ and $3N$ interactions, respectively. The Hamiltonian, initially expressed in terms of nucleon momentum and interactions, is reformulated using normal-ordered operators, predominantly truncated at the two-body level. This reformulation ingeniously incorporates the $3N$ interaction effects within the two-body framework, streamlining the computational complexity.

In this study, we employ two sets of $NN+3N$ chiral forces, e.g., EM1.8/2.0~\cite{PhysRevC.83.031301, PhysRevC.93.011302} and  $NN+3N$(lnl) interactions~\cite{PhysRevC.101.014318}. 
The EM1.8/2.0 interaction is composed of a next-to-next-to-next-to-leading order (N$^3$LO) $NN$ interaction~\cite{PhysRevC.68.041001} soften by the similarity renormalization group (SRG) evolution with momentum resolution scale $\lambda =$ 1.8 fm$^{-1}$ and a next-to-next-to-leading order (N$^2$LO) nonrenormalized three-nucleon forces (3NF) with momentum cutoff $\Lambda =$ 2.0 fm$^{-1}$, where the 3NF is fit to the $^3$H binding energy and the ${^4}$He charge radius based on the soften SRG $\lambda =$ 1.8 fm$^{-1}$ $NN$ interaction~\cite{PhysRevC.83.031301, PhysRevC.93.011302}.
The $NN+3N$(lnl) interaction incorporates both local and nonlocal $3N$ regulators (lnl) and refits $3N$ parameters for $A = 2, 3, 4$ nuclei based on the bare N$^3$LO $NN$ potential~\cite{PhysRevC.101.014318}.
In the present work, we adopt a large SRG scale of $\lambda =$ 2.6 fm$^{-1}$ for the $NN$ interaction~\cite{PhysRevC.68.041001}, and the induced 3NF are neglected.
For the $3N$ part of the $NN+3N$(lnl) interaction, the bare interaction is adopted in the real calculations.
These interactions are acclaimed for their accurate reproduction of the ground state energies and prediction of drip lines in light to medium mass nuclei, which have been extensively validated across various isotopic chains~\cite{PhysRevC.96.014303, PhysRevC.93.011302, Hagen2016, GarciaRuiz2016}. Our computational analysis was executed within a spherical harmonic-oscillator basis including up to $e_{\rm max}= 2n + l = 14$ and limiting the three-body matrix elements to $E_{3\rm max} = 14$. The oscillator frequency of $\hbar w$ is fixed to be 16 MeV.

The VS-IMSRG aims at decoupling the normal-ordered Hamiltonian from the large Hilbert space to a small valence space. This is achieved by solving the flow equation,
\begin{equation}
\frac{dH(s)}{ds}=[\eta(s),H(s)],
\label{FE}
\end{equation}
with an anti-Hermitian generator,
\begin{equation}
\eta(s)\equiv\frac{dU(s)}{ds}U^{\dagger}(s)=-\eta^{\dagger}(s).
\end{equation}

The Magnus formulation of the VS-IMSRG~\cite{PhysRevC.92.034331, HERGERT2016165} is adopted to construct an approximate unitary transformation to first decouple the $^{28}$O core. Subsequently, we construct a valence-space Hamiltonian for valence protons in the $sd$ shell and valence neutrons in the $pf$ shell, effectively reducing the complexity of the full $A$-body problem. 
In our previous work~\cite{YUAN2024138331}, we employed the valence space Hamiltonian to study the deformations in $^{40}$Mg, $^{42}$Si, and $^{44}$S $N=28$ isotones, which yielded good agreement with experimental data.
This approach is refined using the ensemble normal-ordering (ENO) technique detailed in Ref.~\cite{PhysRevLett.118.032502}.
The VS-IMSRG code of Ref.~\cite{imsrg_code} is utilized for that matter.
In practical calculations, the Magnus formalism is employed with all operators truncated at the two-body level~\cite{PhysRevC.92.034331}.
The culmination of this process involves employing the KSHELL shell-model code~\cite{SHIMIZU2019372} for the final diagonalization for valence space Hamiltonians.

\section{Results}

\begin{figure}
    \centering
    \includegraphics[width=0.4\paperwidth]{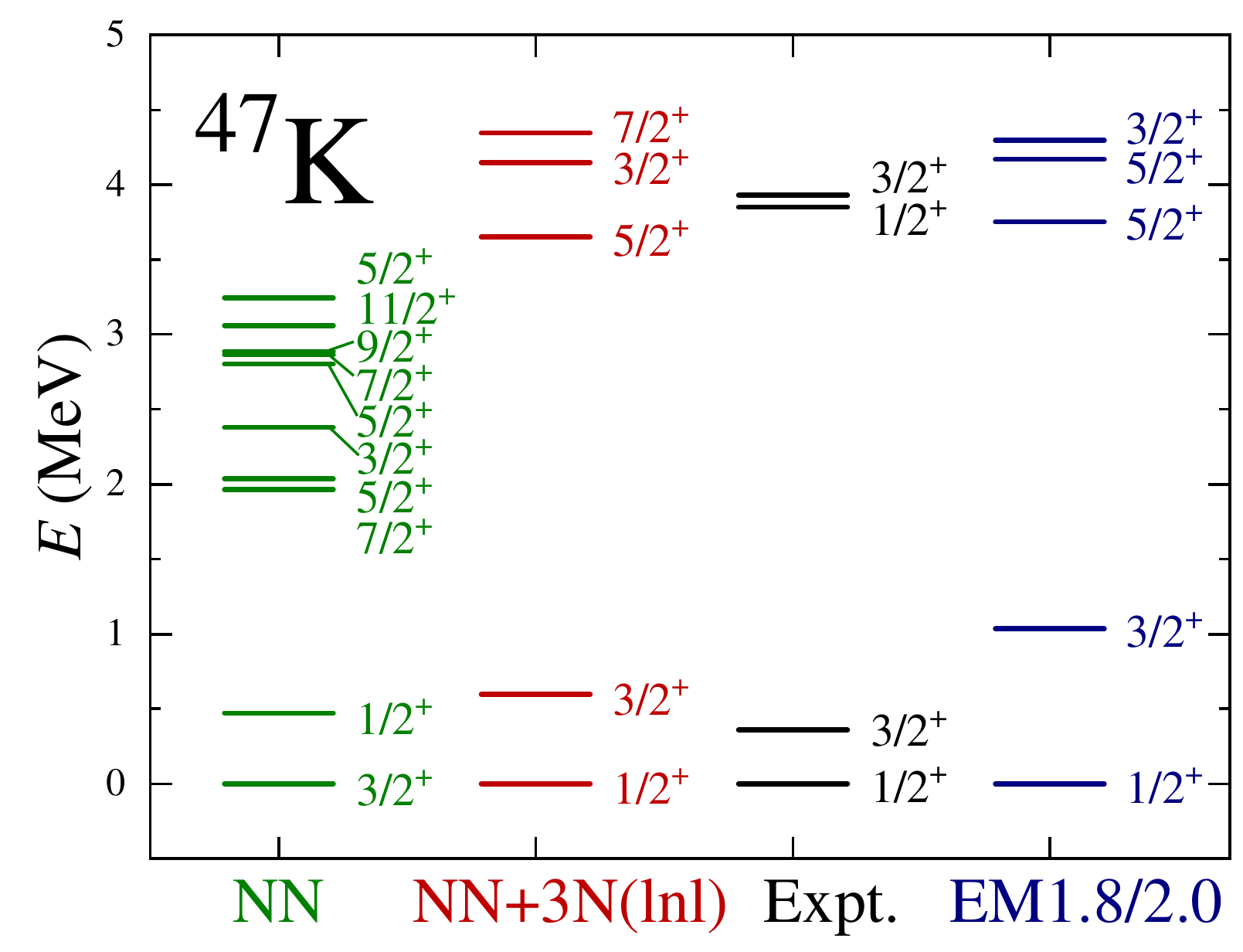}
    \caption{The calculated spectra of $^{47}$K with and  without 3NF for $NN+3N$ (lnl) interaction, along with experimental data~\cite{OGILVIE1987445} and results from EM1.8/2.0 $NN+3N$ interaction.}
    \label{47K}
\end{figure}

Spectroscopic studies of odd-$A$ and odd-odd nuclei play a critical role in validating single-particle structures and refining theoretical models. Neutron-rich P, Cl, and K isotopes, featuring proton holes in $sd$ shell, are of particular interest for elucidating single-particle structures in this notable region of the nuclide chart.
Our study begins with calculations of the energy spectrum for odd-$A$ nuclear isotopes with VS-IMSRG using two sets of $NN+3N$ chiral interaction, i.e., $NN+3N$(lnl) and EM1.8/2.0 interactions. 
Before discussing the results of energy spectrum calculations, we initially focused on assessing the effect of 3NF on the low-lying energy spectra by using $^{47}$K as an example. We compared the results of calculations using the $NN+3N$(lnl) interaction without the inclusion of 3NF (labeled NN), the full $NN+3N$(lnl) interaction, and the EM1.8/2.0 interaction, along with the available experimental data~\cite{OGILVIE1987445}. 
The results are presented in Fig.~\ref{47K}. 
The calculated ground state without including 3NF exhibits significant discrepancies compared to experimental data. The incorporation of 3NF into the VS-IMSRG calculations markedly enhances the consistency of the calculated results with experimental data. This comparison unambiguously underscores the importance of including 3NF in energy spectrum calculations and validates the significant contribution of considering 3NF towards enhancing the accuracy of spectral calculations.

\begin{figure*}[!htb]
\includegraphics[width=0.8\paperwidth]{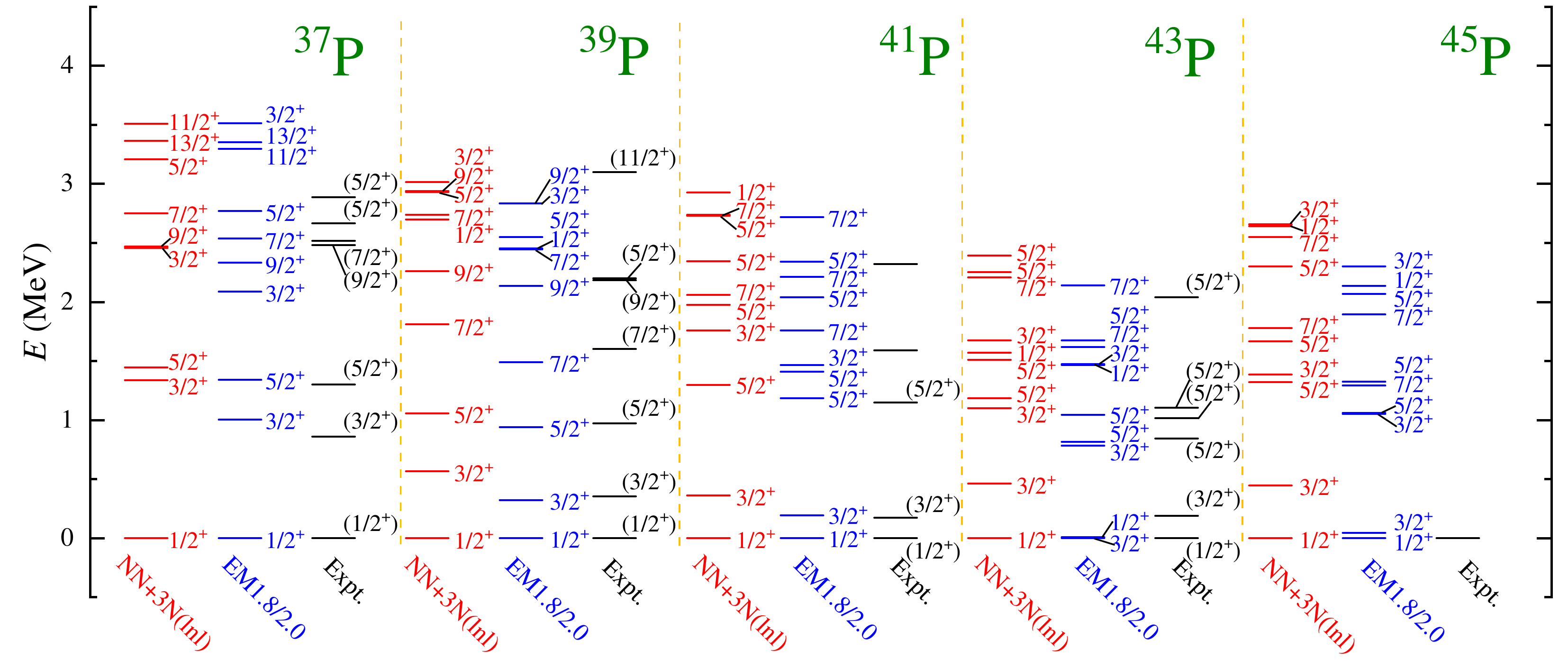}
\caption{The calculated spectra of the odd-$A$ P isotopes with \textit{ab initio} VS-IMSRG using chiral $NN+3N$(lnl) and EM1.8/2.0 interactions, along with available experimental data~\cite{ensdf,PhysRevC.104.014305}.}
\label{P-odd}
\end{figure*}

\begin{figure*}[!htb]
\includegraphics[width=0.8\paperwidth]{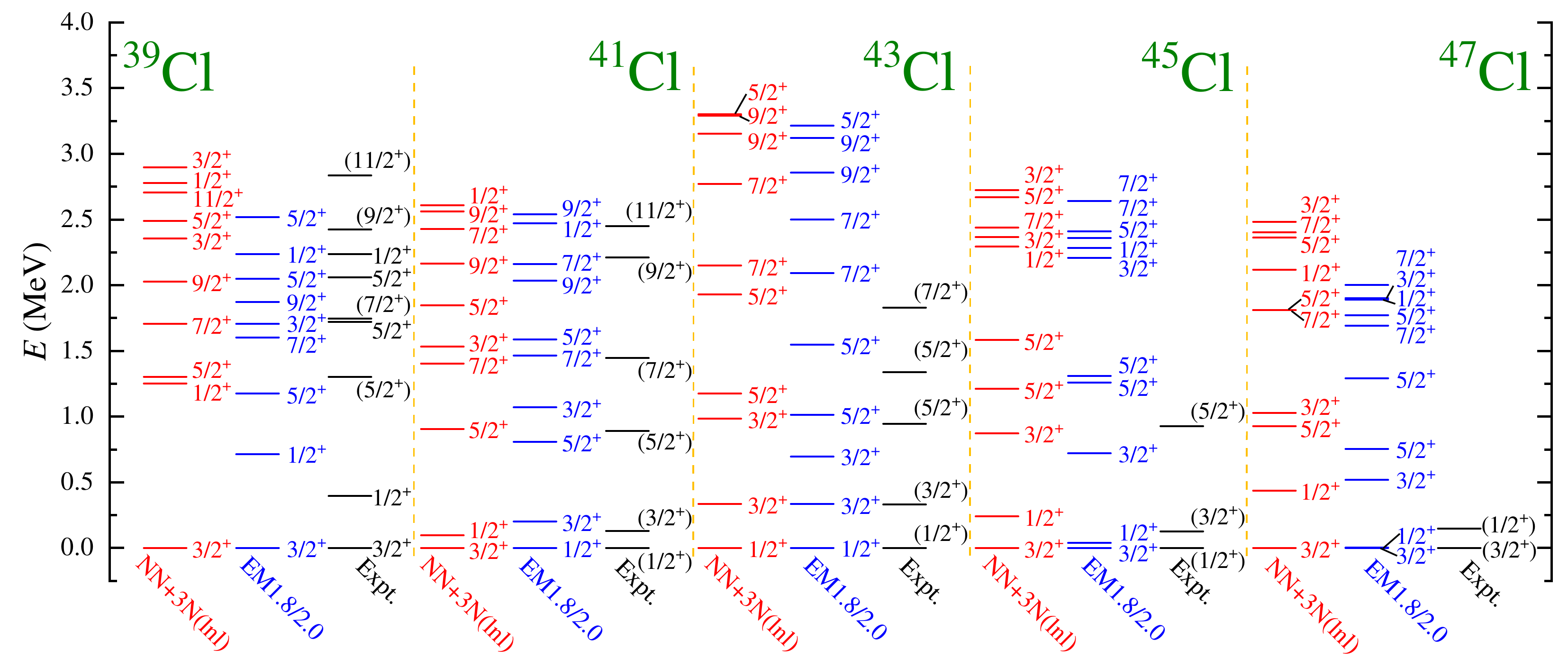}
\caption{Similar to Fig.~\ref{P-odd}, but for neutron-rich odd-$A$ Cl isotopes. Experimental data are taken from Ref.~\cite{ensdf, tripathi2023low, epja/i2004-10019-7, PhysRevC.87.054322}.}
\label{Cl-odd}
\end{figure*}

\begin{figure*}[!htb]
\includegraphics[width=0.8\paperwidth]{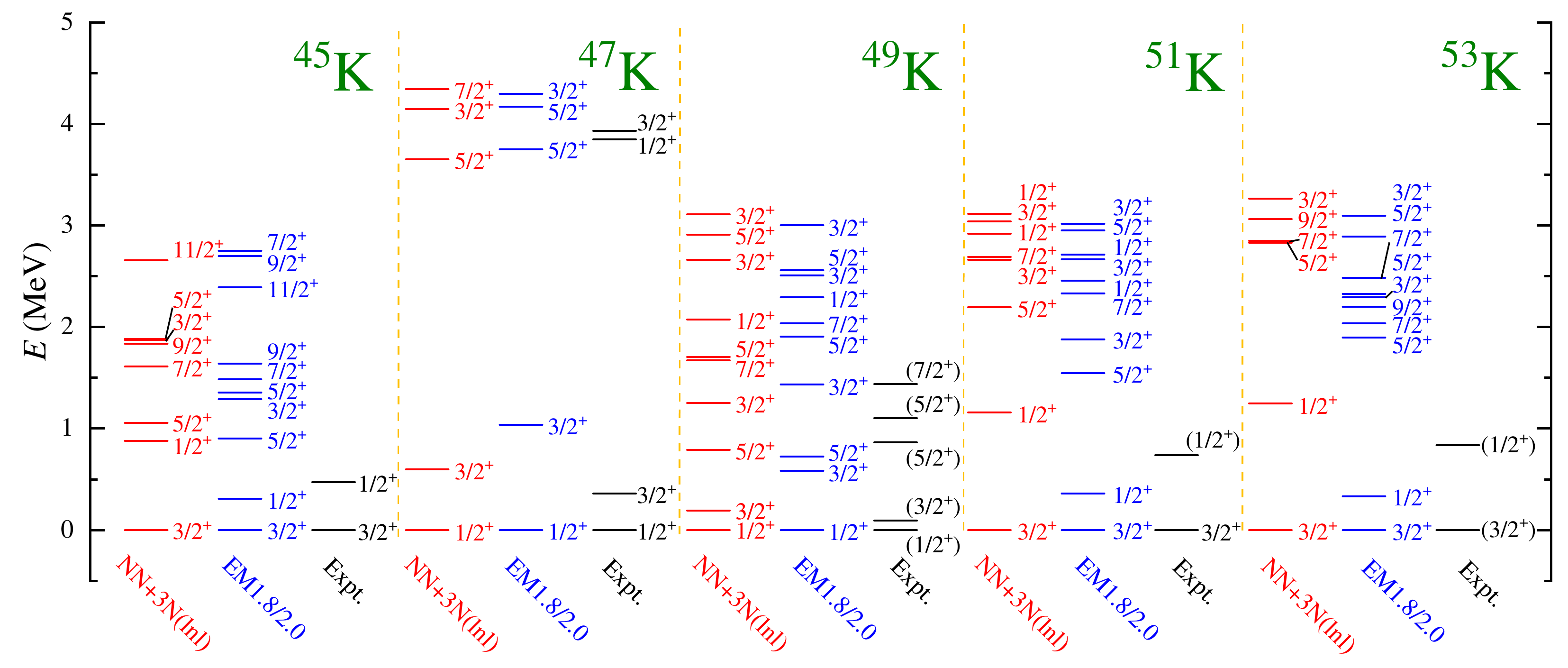}
\caption{Similar to Fig.~\ref{P-odd}, but for neutron-rich odd-A K isotopes. Experimental data are taken from Ref.~\cite{ensdf, PhysRevC.82.034319, SUN2020135215, PhysRevC.4.1621, OGILVIE1987445}.}
\label{K-odd}
\end{figure*}

In the following calculations, we present exclusively the results accounting for the 3NF.
The results of low-lying spectra of neutron-rich  odd-$A$ P, Cl, and K isotopes are illustrated in Figs.~\ref{P-odd}, \ref{Cl-odd}, and \ref{K-odd}, respectively. Moreover, the available experimental data are shown for comparison.

\begin{figure}[!htb]
\includegraphics[width=0.4\paperwidth]{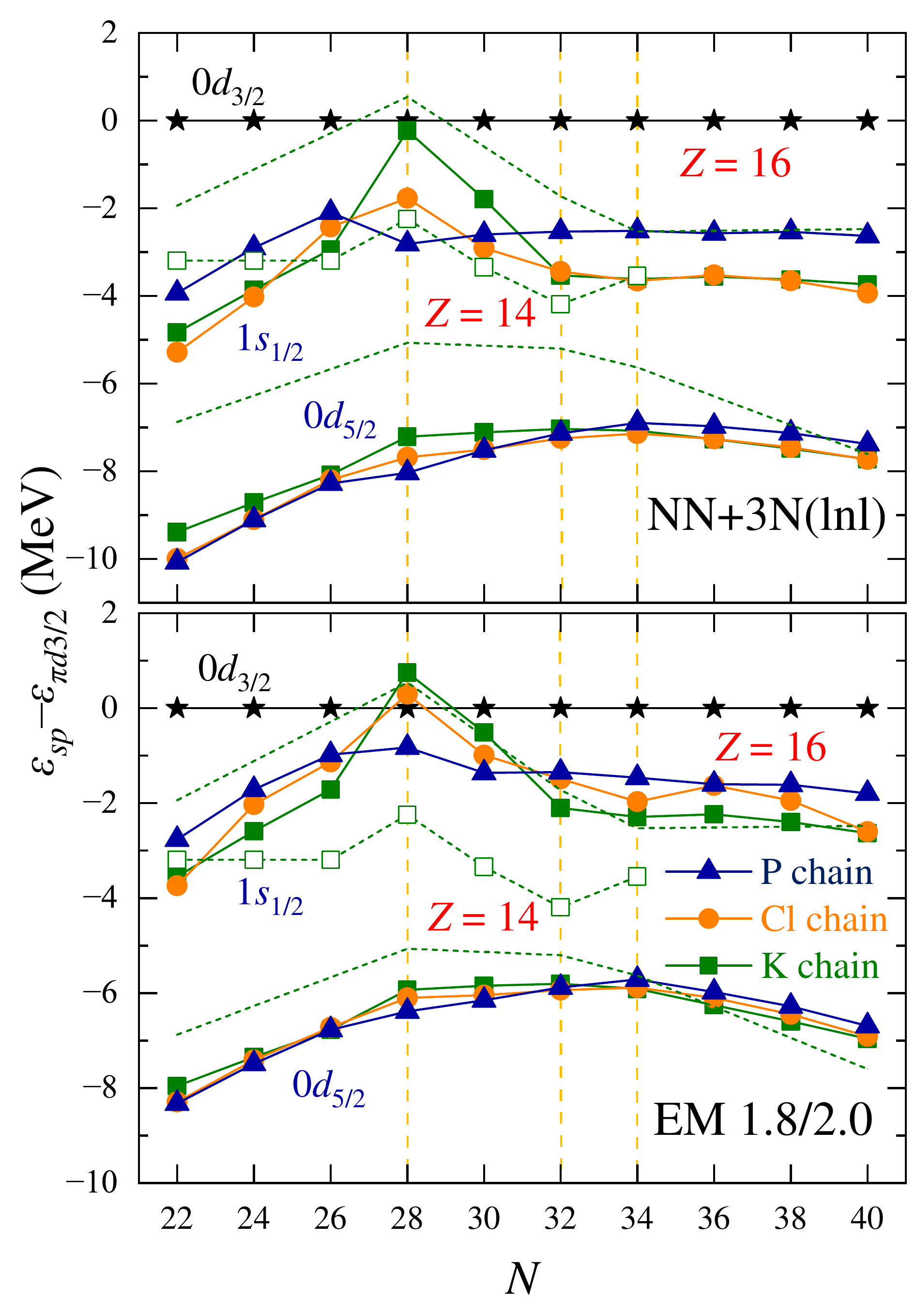}
\caption{The ESPEs of the $\pi 0d_{5/2}$, $\pi 1s_{1/2}$, and $\pi 0d_{3/2}$ orbits in the P, Cl, and K isotopes, for neutron numbers $N=22-40$, were calculated using $NN+3N$(lnl) and EM1.8/2.0 interactions (dashed lines). Furthermore, the calculated ESPEs for K isotopes from  Gorkov-Green’s functions theory (green blank square dotted lines) and the shell model with SDPF-MU interaction (green dotted lines), as reported in Ref.~\cite{PhysRevC.90.034321, physics4010014} are taken for comparison.}
\label{KClP}
\end{figure}

\begin{figure*}
    \centering
    \includegraphics[width=0.75\paperwidth]{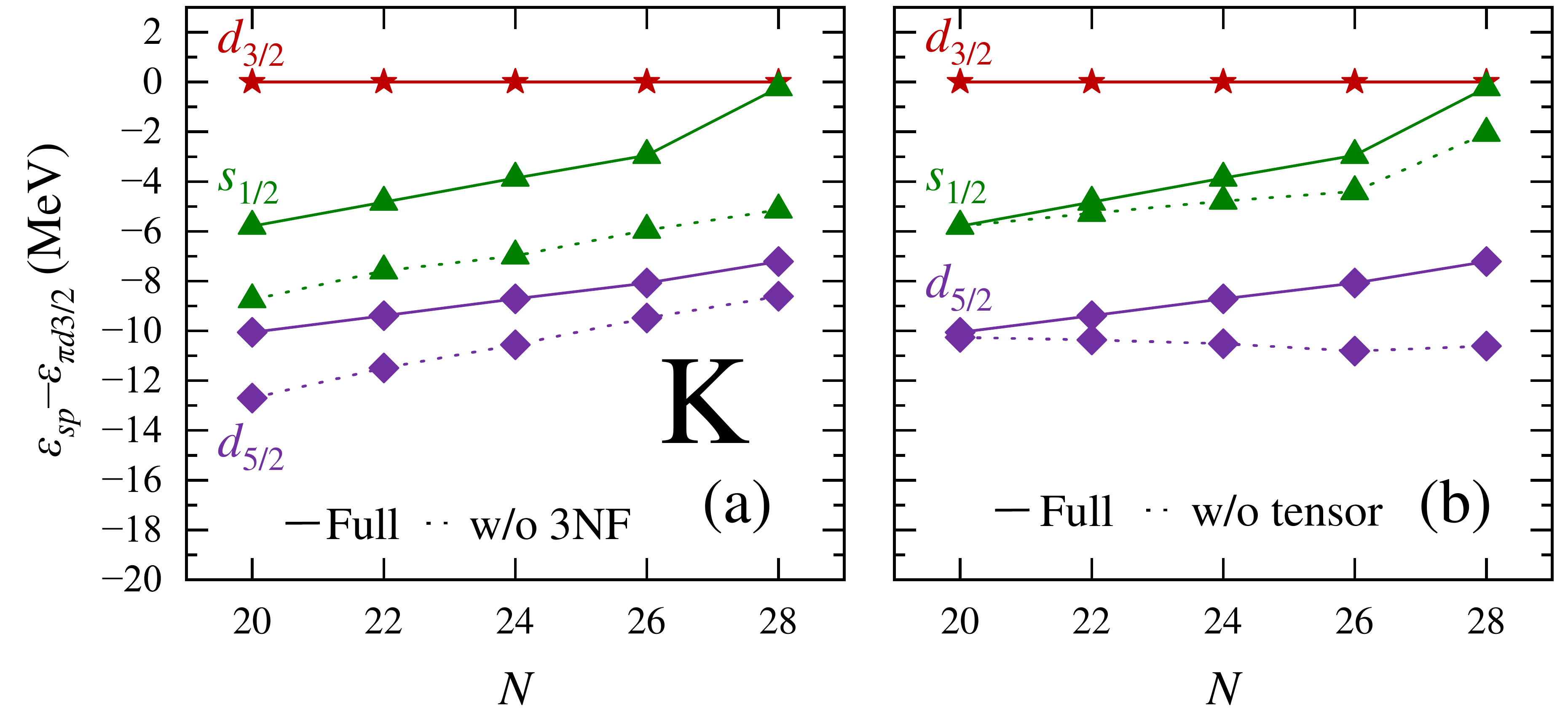}
    \caption{Calculations of ESPEs for K isotope chains utilizing the NN+3N(lnl) interaction are performed. (a) including cases with and without 3NF. (b) the results compare the calculations with and without tensor forces.}
    \label{add}
\end{figure*}

\begin{figure*}[!htb]
\includegraphics[width=0.8\paperwidth]{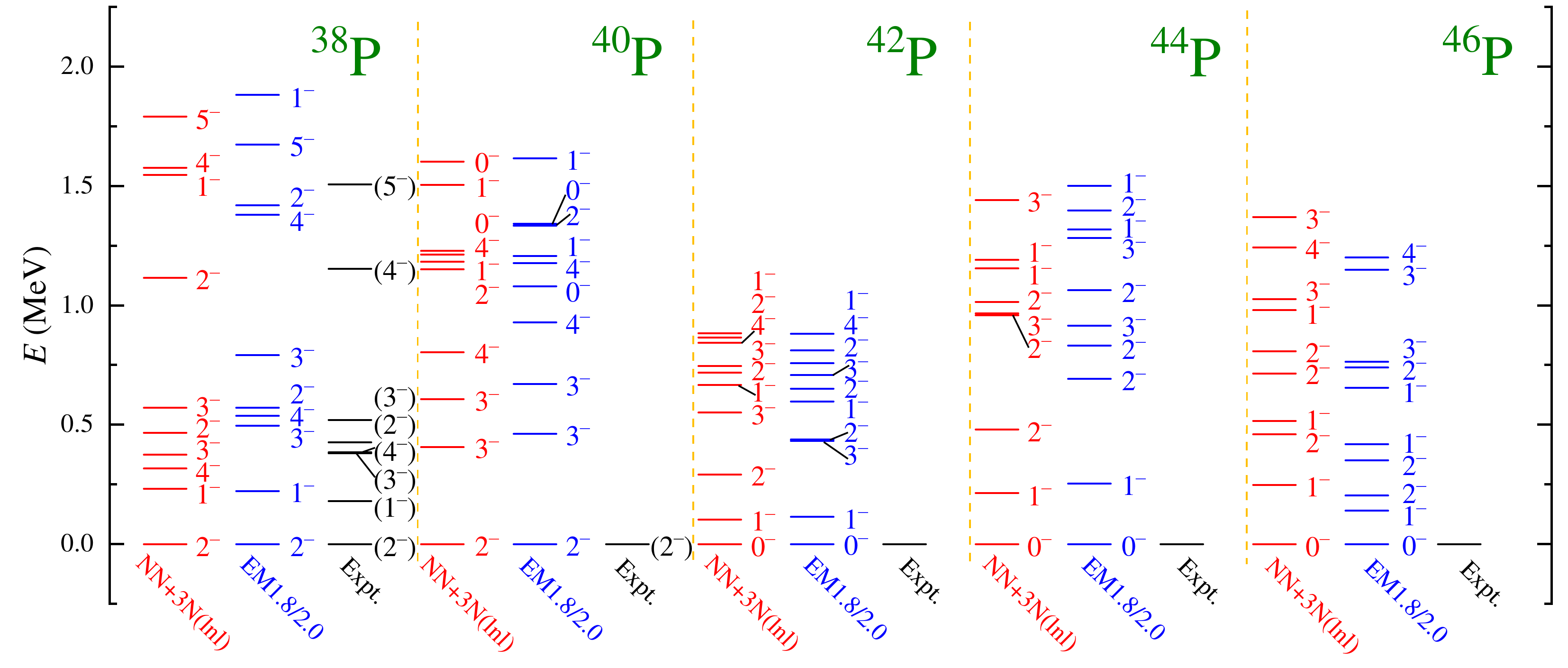}
\caption{The spectra of the even-P isotopes  were calculated using $NN+3N$(lnl) and EM1.8/2.0 interactions, with experimental data taken from Ref.~\cite{ensdf, PhysRevC.104.014305}.}
\label{P-even}
\end{figure*}

\begin{figure*}[!htb]
\includegraphics[width=0.8\paperwidth]{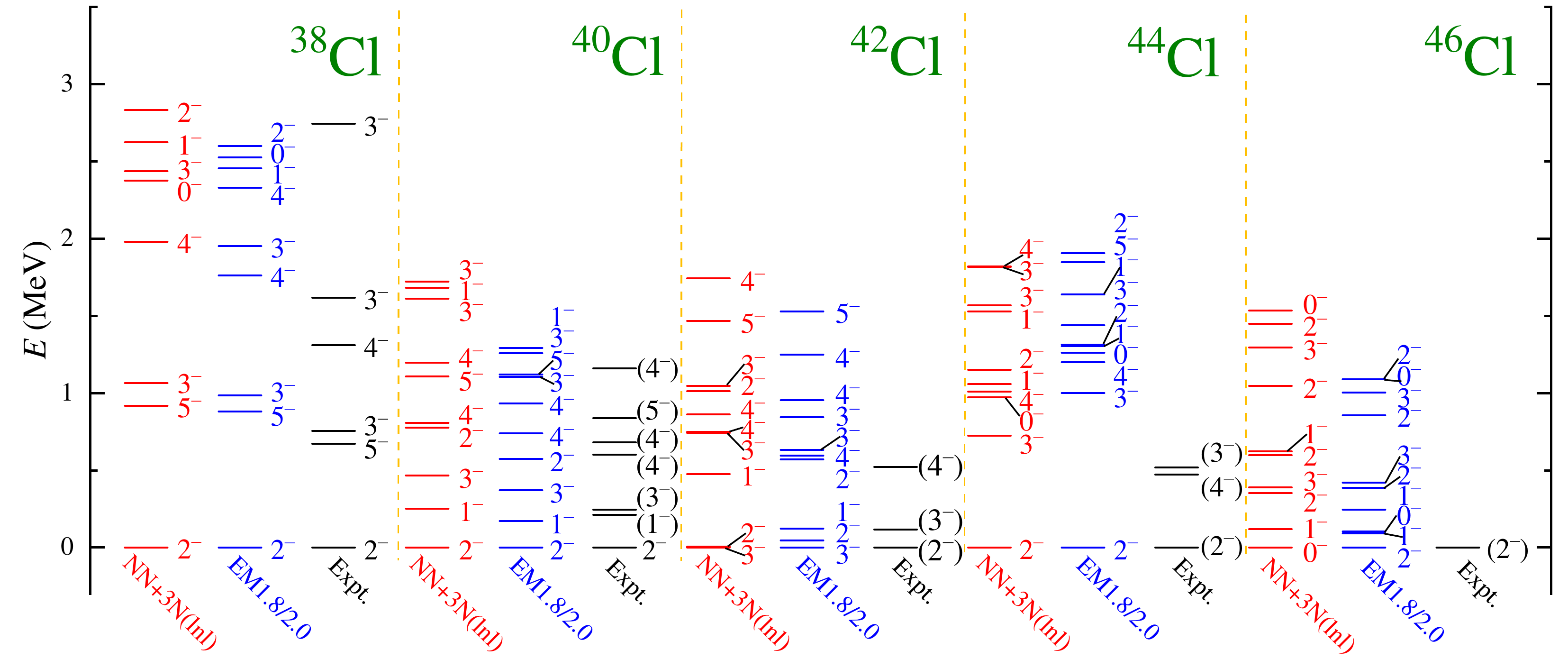}
\caption{Similar to Fig.~\ref{P-even}, but for neutron-rich even-Cl isotopes. Experimental data are taken from Ref.~\cite{ensdf, tripathi2023low,PhysRevC.87.054322, PhysRevC.73.044318}.}
\label{Cl-even}
\end{figure*}

\begin{figure*}[!htb]
\includegraphics[width=0.8\paperwidth]{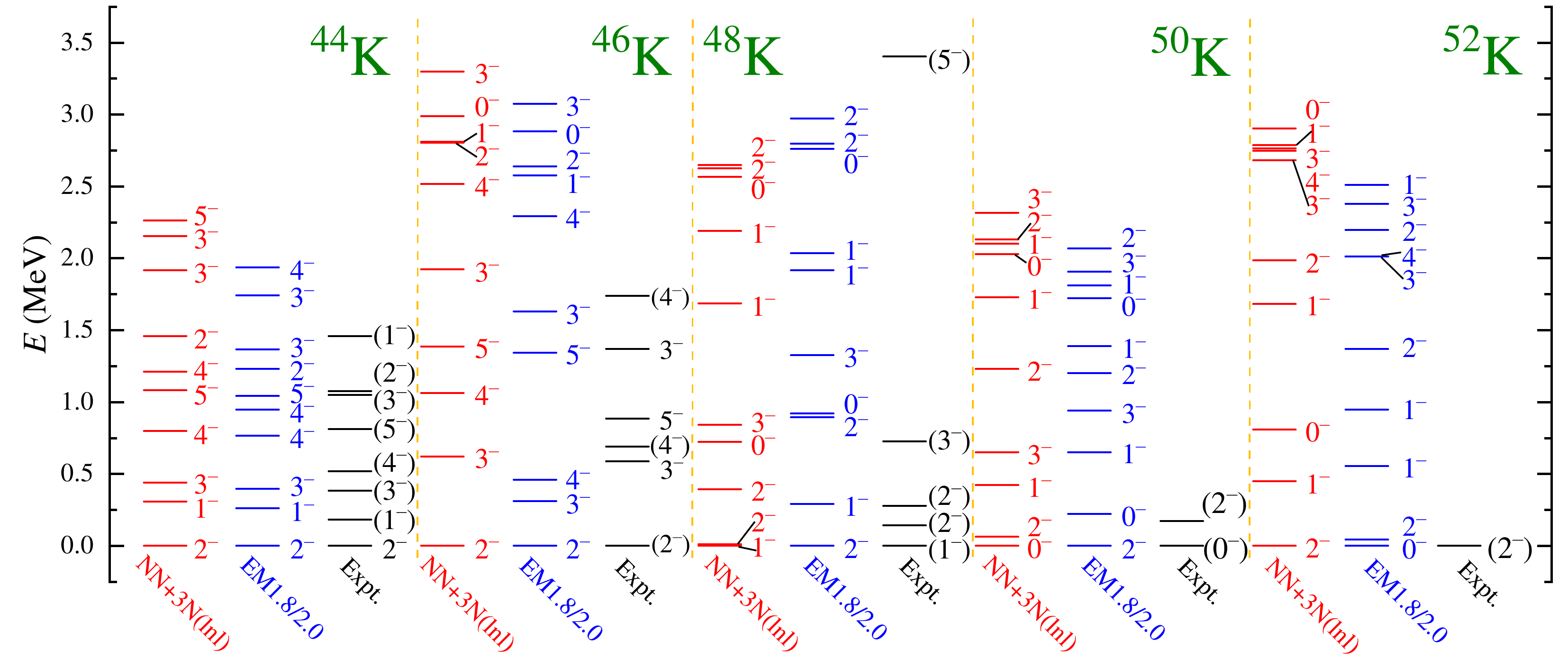}
\caption{Similar to Fig.~\ref{P-even}, but for neutron-rich even-K isotopes. Experimental data are taken from Ref.~\cite{ensdf, PhysRevC.22.427,PhysRevC.84.064301}.}
\label{K-even}
\end{figure*}

The ground state spin and parity of odd-$A$ nuclei could reflect the primary configuration of the last odd nucleon within the shell model framework.
In neutron-rich P isotopes, the shell model traditionally predicts a $1/2^+$ ground state, corresponding to the last odd proton occupying the $\pi 1s_{1/2}$ orbit under normal orbital ordering.
Our \textit{ab initio} VS-IMSRG calculations, utilizing two sets of $NN+3N$ interactions, show good agreements with available experimental data, particularly for the low-lying states of $^{39}$P, where both interactions align closely with experimental data~\cite{PhysRevC.104.014305}.  
Notably, results from the $NN+3N$(lnl) and EM1.8/2.0 interactions are nearly identical, with the notable exception of the low-lying states of $^{43}$P,  where the EM1.8/2.0 interaction predicts a $3/2^+$ ground state, contrary to the experimental suggestion of $1/2^+$~\cite{PhysRevC.104.014305}.
Experimental data on the low-lying states of neutron-rich P isotopes remain sparse, yet both our calculations and available experimental evidence, except for $^{43}$P using the EM1.8/2.0 potential, consistently indicate a $1/2^+$ ground state for odd-A $^{37-45}$P isotopes.
However, the energy level splitting between $E(1/2^+_1)$ and $E(3/2^+_1)$ decreases from $^{37}$P to $^{45}$P. 
Similar results are also obtained in the large-scale shell model calculations~\cite{PhysRevC.74.034322}.
Both two $NN+3N$ interactions predict that the ground state of $^{45}$P is $1/2^+$, while the experimental data is lacking.

Extending the studies to Cl ($Z=17$) isotopes, a notable uncertainty also surrounds the energy spectrum, particularly in isotopes such as $^{41,43,45,47}$Cl. Neutron-rich Cl isotopes are of particular interest due to experimental data suggesting that energy levels of $1/2^+_1$ and $3/2^+_1$ states first reverse with increasing neutron number and then return to the natural order.
As illustrated in Fig.~\ref{Cl-odd}, the prediction of the energy order of the low excited states of the Cl isotopes by both two interactions is also almost identical, except for $^{41}$Cl. 
Note that $^{39}$Cl with $N=22$ displays the expected spin-parity with a $3/2^+$ ground state. However, the situation changes in $^{41}$Cl, where a $1/2^+$ ground state is favored~\cite{PhysRevC.87.054322, PhysRevC.66.037301, PhysRevC.67.024302}. 
This inference draws on observed decay patterns in lighter isotopes, where the yrast $5/2^+_1$ level predominantly decays to the $3/2^+_1$ state, seldom to the $1/2^+_1$ level. Support for a $1/2^+_1$ assignment in $^{41}$Cl also comes from the $\beta$-decay of $^{41}$S~\cite{Winger1998LowEnergyNS}.
The results align with the VS-IMSRG calculations with EM1.8/2.0 interaction.
In $^{45}$Cl, experimental data of Ref.~\cite{Sorlin2004} have suggested that the ground state is $1/2^+_1$, while recent studies suggest a shift towards a $3/2^+_1$ ground state~\cite{tripathi2023low}. The new results with $3/2^+_1$ ground state for $^{45}$Cl are consistent with our VS-IMSRG calculations with both sets of $NN+3N$ interaction.
Similarly, for $^{43}$Cl, robust evidence from in-beam $\gamma$ spectroscopy following fragmentation strongly indicates an inverted $1/2^+_1$ ground state~\cite{Sorlin2004}, while also aligns with VS-IMSRG predictions. Moreover, our VS-IMSRG calculations well reproduce the experimental $3/2_1^+$ and $1/2_1^+$ energy splitting gaps.
It is noted that the energy level order of $^{43}$Cl predicted by VS-IMSRG is in perfect agreement with experimental data.
In addition, VS-IMSRG predicted a $3/2^+_1$ ground state for $^{47}$Cl, consistent with experiments in the Ref.~\cite{tripathi2023low}.

Neutron-rich K isotopes have received considerable attention both experimentally and theoretically~\cite{PhysRevC.82.034319, PhysRevC.74.014313, PhysRevC.21.712, PhysRevC.70.024304}. For the K isotopes, the calculations with the two interactions, $NN+3N$(lnl) and EM1.8/2.0, give the energy level of low-lying states very close, and the results are also in good agreement with experimental data.
Such as the predicted excitation energies for $5/2_1^+$ state for $^{49}$K well align with experimental evidence.
Furthermore, a potential undetected $3/2_2^+$ state lying above it is suggested in our \textit{ab initio} calculations, which is missed in experiments. 
The inversion and restoration of the ground state spin-parity also occur in K isotopes as the neutron number increases up to $N=40$. 
Experiments have confirmed that the ground states of $^{47,49}$K is $1/2^+$~\cite{TOUCHARD1982169, PhysRevLett.110.172503, PhysRevC.82.034319}, while it changes to $3/2^+$ states in nearby odd-$A$ K isotopes.
Our \textit{ab initio} VS-IMSRG calculations using both $NN+3N$(lnl) and EM1.8/2.0 interactions align well with the experimental data, especially the ground state spin-parities and energy splittings of the $3/2_1^+$ and $1/2_1^+$ for the K isotopes. 
For the ground state of $^{49}$K, the large-scale shell model calculations with SDPF-NR interaction predict it is $1/2^+$, while the calculations with SDPF-U and SDPF-MU interactions predict a $3/2^+$ ground state~\cite{PhysRevLett.110.172503}.
Our \textit{ab initio} calculations with both sets of $NN+3N$ interactions give that the ground state of $^{49}$K is $1/2^+$.
Additionally, higher excited states are also predicted by our calculations for the neutron-rich odd-$A$ P, Cl, and K isotopes, which will offer insights for future experimental endeavors.

The variation of ESPEs as a function of neutron number $N$ or proton number $Z$ is crucial for understanding shell evolution, as emphasized in Ref.~\cite{RevModPhys.92.015002, PhysRevLett.95.232502}. 
Shell model calculations have highlighted the importance of central and tensor forces in altering the $N=20$ and $N=28$ shell gaps, particularly noting the significant impact of tensor forces on the energy shift of spin-orbit partners \cite{PhysRevC.86.034314, RevModPhys.92.015002}.
The evolution for the  $3/2_1^+$ and $1/2_1^+$ states in neutron-rich odd-$A$ P, Cl, and K isotopes are of particular interest for the nuclear structure, which is attributed to the evolution of $\pi 0d_{3/2}$ and $\pi 1s_{1/2}$ single-particle states. To extend the understanding of shell evolution in neutron-rich P, Cl, and K isotopes, the ESPEs for the proton single particle state are comprehensively analyzed based on the effective valence-space Hamiltonian derived from the VS-IMSRG.

The calculated ESPEs are depicted in Fig.~\ref{KClP}. The energy gap of $\pi 0d_{5/2}$ and $\pi 1s_{1/2}$ relative to $\pi 0d_{3/2}$ single-particle states are presented, which could reveal significant shifts in orbital energy splitting as neutron number $N$ increases. 
Specifically, neutron-rich P, Cl, and K isotopes exhibit nearly identical variations in the orbital gaps between $\pi 0d_{3/2}$ and $\pi 0d_{5/2}$ across the neutron numbers $N=22-40$. 
Furthermore, the gaps between $\pi 1s_{1/2}$ and $\pi 0d_{3/2}$ orbits display a similar pattern in neutron-rich P, Cl, and K isotopes. They initially decrease and then increase with the addition of neutrons, stabilizing once $N \geq 32$. 
Shell model calculations have suggested that the decreasing of the shell gap between $\pi 1s_{1/2}$ and $\pi 0d_{3/2}$ is mainly affected by the tensor force~\cite{SMIRNOVA2010109}.
It is important to note that around $N=28$, the energies of $\pi 0d_{3/2}$ and $\pi 1s_{1/2}$ orbits are nearly degenerate, which form a pseudo-SU3 doublet that enhances the quadrupole correlations of the configurations with open-shell neutrons~\cite{PhysRevC.79.014310}. Notably, beyond $N=28$, the shell gap increases, regaining its standard order. 
Furthermore, our VS-IMSRG calculations are also compared with the results from Gorkov-Green's functions theory~\cite{PhysRevC.90.034321} and shell model calculations with SDPF-MU interaction~\cite{physics4010014}.
In Gorkov-Green’s functions theory, ESPEs of $\pi 0d_{3/2}$ and $\pi 1s_{1/2}$ recall the fragmented $3/2^+$ and $1/2^+$ strengths derived from one-proton addition and removal processes on neighboring Ca isotones~\cite{PhysRevC.84.064317}. This comparison unveils a consistency in the evolutionary trends of ESPEs predicted by the VS-IMSRG and Gorkov-Green's function theory, especially beyond the neutron number $N=26$.  Compared to the shell model results, derived from the Ca isotope chain, both methods indicate the disappearance of the magic number $Z=16$ at $N=28$. Notably, beyond $N=28$, the shell model results closely align with those obtained using the EM1.8/2.0 interaction.

It is noteworthy that the EM1.8/2.0 interaction gives $\pi 0d_{3/2}$ and $\pi 1s_{1/2}$ orbits that are inverted at $N=28$, with a crossover in the ESPEs in neutron-rich Cl and K isotopes. However, the VS-IMSRG predicts the $^{45}$Cl ground state to be $3/2^+$. Although ESPE provides a preliminary framework for nuclear energy levels, the interactions between nucleons and the collective effects arising from them play a decisive role in the actual occupation and formation of the ground state of the nucleus~\cite{PhysRevC.86.034314}.  Interestingly, although the $NN+3N$(lnl) calculations do not show a crossover between the $\pi 1s_{1/2}$ and $\pi 0d_{3/2}$ orbits for neutron-rich P, Cl, and K isotopes, the VS-IMSRG calculations correctly predict the experimental energy level inversion at $^{47}$K and $^{51}$K for K isotopes, and at $^{41}$Cl for Cl isotopes.
Notably, the calculated ESPEs clearly show the evolution of proton shells, in which the $Z=14$ sub-shell exists while the $Z=16$ disappearances in the neutron-rich P, Cl, and K isotopes in the vicinity of $N=28$.

Addressing 3NF, this analysis illuminates their contributions to shell evolution. Fig.~\ref{add}(a) depicts the ESPE of the K isotope chain for $N=20-28$ with and without 3NF for the $NN+3N$(lnl) interaction. With the exclusive consideration of $NN$ interaction, a pronounced energy gap exists between the $0d_{3/2}$ and $1s_{1/2}$ orbits, preserving the sub-shell at $Z=16$, which contradicts the ground-state level inversion observed in $^{47}$K~\cite{OGILVIE1987445}. This implies that calculations relying solely on $NN$ interaction are insufficient to account for the shell evolution, especially for the disappearance of the magic number at $Z=16$. Conversely, incorporating 3NF addresses this shortfall by elevating the ESPEs. With an increase in the neutron number, the energy gap at $Z=16$ diminishes, establishing the groundwork for the disappearance of the shell structure at $N=28$ and underscoring the importance of 3NF in describing shell evolution. Moreover, through the application of the spin-tensor decomposition method~\cite{PhysRevC.86.034314, SMIRNOVA2010109, PhysRevLett.112.042502}, it becomes possible to isolate different components of the effective $NN$ interaction, facilitating a qualitative analysis of contributions of specific components to the evolution of ESPE.
Fig.~\ref{add}(b) compares the calculated ESPEs with and without considering tensor forces based on the effective Hamiltonian derived from $NN+3N$(lnl) interaction. The $0d_{5/2}$ orbit is more significantly impacted by tensor forces compared to the $1s_{1/2}$ orbit, highlighting the substantial influence of tensor forces on spin-orbit splitting.  When tensor forces are considered, the $0d_{3/2}$ and $1s_{1/2}$ orbits become almost degenerate at $N=28$, with an energy gap of only 0.228 MeV, suggesting the disappearance of the magic number at $Z=16$ and aligning with the energy spectra results. The omission of tensor forces results in an increased shell gap between the $0d_{5/2}$ and $1s_{1/2}$ orbits to 2 MeV, emphasizing the influence of tensor forces on shell evolution.
Similar situations also occur in the P and Cl isotopes, in which the 3NF and tensor forces play important roles in shell evolutions.

Furthermore, 
a marked decrease in the energy gap of the $1s_{1/2}$ and $0d_{3/2}$ orbits at $N=28$ is observed in the K isotopes, shown in Fig.~\ref{add}(a), which is absent in the same calculations with only $NN$ interaction. Moreover, the influence of tensor forces for the marked decrease in the energy gap is considered negligible, as demonstrated in Fig.~\ref{add}(b). To further understand the mechanism, we calculate the average occupations of valence nucleons for the K isotopes with and without 3N forces, as presented in Table~\ref{occ}. These results show that as the neutron number increases, the valence neutrons sequentially fill the $\nu 0f_{7/2}$ orbit in K isotopes. Notably, at $N=28$, a sudden increase occurs in the occupancy of $\pi 0d_{3/2}$ orbit, accompanied by a concurrent decrease in occupancy for the $\pi 1s_{1/2}$ orbit. The attraction of the $\nu 0f_{7/2}$ orbit towards both the $\pi 0d_{3/2}$ and $\pi 1s_{1/2}$ orbits implies that the enhanced occupancy of the $\pi 0d_{3/2}$ orbit results in a more pronounced reduction in its energy compared to the $\pi 1s_{1/2}$ orbit, leading to the degeneracy of the $\pi 0d_{3/2}$ and $\pi 1s_{1/2}$ orbits and resulting the marked decrease in the energy gap of the $1s_{1/2}$ and $0d_{3/2}$.

\begin{table*}[]
\centering
\caption{The calculated average occupancies of valence nucleons for K isotope using the only $NN$ and the $NN+3N$(lnl) interactions.}
\begin{tabular}{cccccccccccccccc}
\hline\hline
        & \multicolumn{2}{c}{$\pi 0d_{5/2}$}& \multicolumn{2}{c}{$\pi 0d_{3/2}$} & \multicolumn{2}{c}{$\pi 1s_{1/2}$} & \multicolumn{2}{c}{$\nu 0f_{7/2}$} & \multicolumn{2}{c}{$\nu 0f_{5/2}$} & \multicolumn{2}{c}{$\nu 1p_{3/2}$} & \multicolumn{2}{c}{$\nu 1p_{1/2}$} \\ 
        &$NN+3N$ & $NN$ & $NN+3N$ & $NN$ & $NN+3N$ & $NN$ & $NN+3N$ & $NN$ & $NN+3N$ & $NN$ & $NN+3N$ & $NN$ & $NN+3N$ & $NN$ \\\hline
$^{41}$K & 5.994 & 5.999 & 3.028 & 3.040 & 1.978 & 1.962 & 1.894 & 1.982 & 0.034 & 0.005 & 0.063 & 0.011 & 0.010 & 0.002 \\
$^{43}$K  & 5.961 & 5.985 & 3.074 & 3.025 & 1.965 & 1.990 & 3.784 & 3.491 & 0.082 & 0.050 & 0.117 & 0.437 & 0.017 & 0.023 \\
$^{45}$K & 5.952 & 5.982 & 3.076 & 3.065 & 1.972 & 1.952 & 5.653 & 5.348 & 0.128 & 0.087 & 0.195 & 0.533 & 0.024 & 0.032 \\
$^{47}$K & 5.973 & 5.992 & 3.777 & 3.118 & 1.250 & 1.890 & 7.485 & 6.942 & 0.112 & 0.115 & 0.377 & 0.886 & 0.027 & 0.058 \\ 
\hline\hline
\end{tabular}
\label{occ}
\end{table*}

The above calculations have shown that the \textit{ab initio} VS-IMSRG approach provides good descriptions for the odd-$A$ nuclear energy spectrum, and the shell evolution can also be well explained with the help of the ESPE based on the effective valence space Hamiltonian derived from VS-IMSRG using the EM1.8/2.0 and $NN+3N$(lnl) interactions. 
Descriptions of odd-odd nuclei have shown that they are more sensitive to interactions~\cite{Sun_2024}.
We extend our analysis to study the energy spectrum of odd-odd nuclei. For P isotopes in Fig.~\ref{P-even}, limited experimental data poses a challenge. Nonetheless, our \textit{ab initio} calculations have shown consistency with the available experimental data in $^{38}$P and $^{40}$P, which are also in line with large-scale shell model calculations with SDPF-MU interaction~\cite{PhysRevC.104.014305}.  No experimental data of low-lying state information of $^{42,44,46}$P has been reported. The predicted level scheme for $^{42,44,46}$P is shown and both interactions suggest a $0^-$ for the ground states, with experimental confirmation required to substantiate these theoretical predictions.
Turn to Cl isotopes in Fig.~\ref{Cl-even}, the VS-IMSRG calculations reproduce the low excitation spectrum of $^{38}$Cl well and hint at the possible presence of an undetected $2_2^-$ state in $^{40}$Cl. 
In the case of $^{42,44,46}$Cl isotopes, experiments have suggested that the ground states of those isotopes are $2^-$~\cite{PhysRevC.73.044318, PhysRevC.87.054322, tripathi2023low}.
VS-IMSRG calculations for the ground state of $^{44}$Cl are consistent with the results in Ref.~\cite{tripathi2023low}, while our VS-IMSRG calculations with both interactions predict that the ground state of $^{42}$Cl is $3^-$. 
For $^{46}$Cl, our \textit{ab initio} calculations with $NN+3N$(lnl) interaction and shell model calculations with EPQQM interaction~\cite{PhysRevC.86.024321} give the ground state being $0^-$, while \textit{ab initio} calculations with EM1.8/2.0 interaction and large-scale shell model calculations with SDPF-U interaction give that the ground state is $2^-$~\cite{PhysRevC.86.024321}. The discrepancy in the predictions for the spin-parity of the $^{46}$Cl ground state underscores the urgent need for definitive experimental evidence.

Focusing on odd-odd K isotopes, as illustrated in Fig.~\ref{K-even}, experimental studies have firmly established the ground state of $^{44}$K as $2^-$~\cite{PhysRevC.18.1803, PhysRevC.21.2135}. This experimental determination aligns with our predictions from both EM1.8/2.0 and $NN+3N$(lnl) interactions, which accurately depict the ground and low-lying excited states. Theoretical calculations suggest that the ground state of $^{46}$K is also $2^-$ with mainly dominated by $\pi (0d_{3/2})^{-1} \nu (0f_{7/2})^{-1}$ configuration, the results are consistent with results in Ref.~\cite{PhysRevC.22.427}. 
The two interactions offer differing predictions for the ground state of $^{48}$K. The predicted ground states from  $NN+3N$(lnl) is $1^-$, which is consistent with the suggested ground states in Ref. \cite{PhysRevC.84.064301}, while the calculations with EM1.8/2.0 give the ground state is $2^-$.
The ground state of $^{50}$K and $^{52}$K remains experimentally unconfirmed, highlighting a critical area for future research to validate our current understanding of nuclear forces.
Moreover, the excited states are also predicted in our \textit{ab initio} VS-IMSRG calculations, which may provide help for future experiments.

\section{Summary}
We employed the \textit{ab initio} VS-IMSRG method, based on two sets of chiral $NN +3N$ interactions, i.e., EM1.8/2.0 and $NN+3N$(lnl) interaction,  to precisely predict and elucidate the shell evolution properties of P, Cl, and K isotopic chains. 
The low-lying spectra of neutron-rich odd-$A$  nuclei P, Cl, and K isotopes were systematically investigated, which has shown that different isotopes exhibit varied phenomena with the increase of neutron number. The ground state spin-parties act as sensitive probes of nuclear wave functions, serving as powerful tools for studying nuclear structures in isotopes far from stability. 
Initially, we demonstrate that the inclusion of three-nucleon forces significantly enhances the accuracy of energy spectrum calculations.
Considering three-nucleon forces, our VS-IMSRG calculations have successfully reproduced the low-lying states for neutron-rich P, Cl, and K isotopes, especially the evolution of $3/2_1^+$ and $1/2_1^+$ states in those isotopes.
Further calculations of ESPEs have confirmed that these energy level shifts are predominantly driven by monopole interactions, leading to a reversal of the $\pi 0d_{3/2}$ and $\pi 1s_{1/2}$ orbits around $N=28$. Both three-nucleon forces and tensor forces play important roles in this process. Interestingly, available data for the P isotopic chain is limited. Our calculations indicate a possible energy level inversion in $^{43}$P at $N=28$, necessitating further investigation.

We also conducted a detailed investigation of the spectral properties of odd-odd nuclei, which has significantly broadened our understanding of these isotopic chains and their behavior in regions far from stability. The results from these calculations show a strong correlation with the limited experimental data, thereby confirming the validity of our computational approach. We also offer predictions for the energy spectra of $^{40,42,44,46}$P, $^{46}$Cl, and $^{52}$K, which would be reached in future experiments.

\textit{Acknowledgments.}~
 This work has been supported by the National Key R\&D Program of China under Grant No. 2023YFA1606403; the National Natural Science Foundation of China under Grant Nos.  12205340, 12175281, 12347106, and 12121005;  the Gansu Natural Science Foundation under Grant No. 22JR5RA123 and 23JRRA614;  the Strategic Priority Research Program of Chinese Academy of Sciences under Grant No. XDB34000000; the Key Research Program of the Chinese Academy of Sciences under Grant No. XDPB15; the State Key Laboratory of Nuclear Physics and Technology, Peking University under Grant No. NPT2020KFY13. The numerical calculations in this paper have been done on Hefei advanced computing center.

\bibliography{Ref}

\end{document}